\documentclass[mypaper,7pt,twoside]{CoAst}
\usepackage{epsf,graphicx,fancyhdr,booktabs,amsmath,amssymb,bm}
\input{CoAst_layo.sty}
\newcommand{\discuss}[2]{{\textbf{\small#1:}}{ \small #2 \normalsize \par}}
\def\rfr{\smallskip\par\noindent
        \hangindent=7truemm
        \hangafter=1}
\def\bu{\bm{u}}
\def\bx{\bm{\xi}}

\def\ob{\overline}

\def\dt{\partial_t}
\def\di{\partial_i}
\def\dj{\partial_j}

\def\kz{\,\delta_{i3}}
\def\ui{u_i}

\def\uj{u_j}

\def\pp{p^\prime}
\def\Tp{T^\prime}
\def\hd{\phantom-\ }
\begin{document}
\sf
\chapterCoAst{The effect of convection on pulsational stability}
             {G\"unter Houdek}
\Authors{\vspace{-3mm}G. Houdek} 
\Address{\vspace{-3mm}Institute of Astronomy, University of Cambridge, 
                      Cambridge CB30HA, UK}
\vspace{-3mm}
\noindent
\begin{abstract}
\vspace{-2mm}
A review on the current state of mode physics in classical pulsators
is presented. Two, currently in use, time-dependent convection models 
are compared and their applications on mode stability are discussed 
with particular emphasis on the location of the Delta Scuti 
instability strip.
\end{abstract}

\vspace{-2mm}
\section{Introduction}
\vspace{-2mm}
Stars with relatively low surface temperatures show distinctive 
envelope convection zones which affect mode stability. Among the
first problems of this nature was the modelling %of the location
of the red edge of the classical instability strip (IS) in the 
Hertzsprung-Russell (H-R) diagram. The first pulsation calculations 
of classical pulsators without any pulsation-convection modelling 
predicted red edges which were much too cool and which were at 
best only neutrally stable. What follows were several attempts 
to bring the theoretically predicted location of the red edge in 
better agreement with the observed location by using time-dependent 
convection models in the pulsation analyses (Dupree 1977; 
Baker \& Gough 1979; Gonzi 1982; Stellingwerf 1984). More
recently several authors, e.g. Bono et al. (1995, 1999), Houdek (1997, 2000), 
Xiong \& Deng (2001, 2007), Dupret et al. (2005)
were successful to model the red edge of the classical IS.
\begin{figure}
\centering
\mbox{
\begin{minipage}[ht!]{4.3cm}
\includegraphics[width=\textwidth]{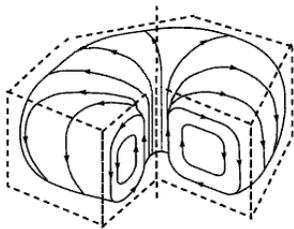}
\end{minipage}
\hspace{0.1cm}
\begin{minipage}[ht!]{6.62cm}
\vspace{-0.2cm}
\caption{Sketch of an overturning hexagonal (dashed lines) 
convective cell. Near the centre the gas raises from the 
hot bottom to the cooler top (surface) where it moves nearly 
horizontally towards the edges, thereby loosing heat. The 
cooled gas then descends along the edges to close the circular
flow. Arrows indicate the direction of the flow pattern.
}
\end{minipage}
}%\mbox
\label{fig:1}
\end{figure}
These authors report, however, that different physical mechanisms 
are responsible for the return to stability. For example,
Bono et al. (1995) and Dupret et al. (2005) report 
that it is mainly the convective heat flux, Xiong \& Deng (2001)
the turbulent viscosity, and Baker \& Gough (1979) and 
Houdek (2000) predominantly the momentum flux (turbulent pressure $p_{\rm t}$)
that stabilizes the pulsation modes at the red edge.
\begin{table}
\centering
\caption{Summary of time-dependent convection model differences.}
\vspace{1em}
\begin{tabular}[ht]{ll}
%\hline
Balance between buoyancy \&        &Kinetic theory of accelerating\\
turbulent drag (Unno 1967, 1977)   &eddies (Gough 1965, 1977a)\\
\hline
\addlinespace[3pt]
- acceleration terms of convective &- acceleration terms included: $w$,\\
\hd fluctuations $w, \Tp$ neglected&\hd $\Tp$ evolve with growth rate $\sigma$\\
\addlinespace[3pt]
- nonlinear terms approximated            &- nonlinear terms are neglected\\  
\hd by spatial gradients $\propto1/\ell$  &\hd during eddy growth\\  
\addlinespace[3pt]
- $\di p^\prime$ neglected in momentum equ.&- $\di p^\prime$ included in Eq. (1)\\
\addlinespace[3pt]
- characteristic eddy lifetime:    &- $\tau=2/\sigma$ determined stochas-\\
\hd $\tau\simeq\ell/2w$            &\hd tically from parametrized shear\\
                                   &\hd instability\\
\addlinespace[3pt]
- variation $\ell_1=\delta\ell/\ell$ (Unno 1967):&- variation of mixing length\\
\hd $\omega\tau<1:\ \ell_1\sim H_1$&\hd according to rapid distortion\\
\hd $\omega\tau>1:\ \ell_1\sim r_1$&\hd theory (Townsend 1976), i.e. \\
\hd or (Unno 1977):                &\hd variation also of eddy shape\\
\hd $\ell_1\sim (1+{\rm i}\omega^2\tau^2)^{-1}(H_1-{\rm i}\omega\tau\rho_1/3)$\\
\hd ($H$ is pressure scale height) &\\
\addlinespace[3pt]
- turbulent pressure $p_{\rm t}$ neglected &- $p_{\rm t}=\ob{\rho}$\,$\ob{ww}$ included in mean\\
\hd in hydrostatic support equation        &\hd equ. for hydrostatic support\\
\addlinespace[3pt]
\hline
\end{tabular}
\label{tab:tc_comp}
\end{table}
\vspace{-3pt}
\section{Time-dependent convection models}
\vspace{-4pt}
The authors mentioned in the previous section used different implementations
for modelling the interaction of the turbulent velocity field 
with the pulsation. In the past various time-dependent convection models 
were proposed, for example, by Schatzman (1956), Gough (1965, 1977a), 
Unno (1967, 1977), Xiong (1977, 1989), Stellingwerf (1982), Kuhfu\ss\ (1986), 
Canuto (1992), Gabriel (1996), Grigahc\`ene et al. (2005).
Here I shall briefly review and compare the basic concepts of two, currently in
use, convection models. The first model is that by Gough (1977a,b), which 
has been used, for example, by Baker \& Gough (1979), Balmforth (1992) 
and by Houdek (2000). The second model is that by Unno (1967, 1977), 
upon which the generalized models by Gabriel (1996) and Grigahc\`ene et al. 
(2005) are based, with applications by Dupret et al. (2005).\\
Nearly all of the time-dependent convection models assume the Boussinesq 
approximation to the equations of motion. The Boussinesq approximation relies
on the fact that the height of the fluid layer is small compared with 
the density scale height. It is based on a careful scaling argument and 
an expansion in small parameters (Spiegel \& Veronis 1960; Gough 1969). 
The fluctuating convection equations for an inviscid Boussinesq fluid in 
a static plane-parallel atmosphere are 
\begin{eqnarray}
\vspace{-3pt}
\dt\ui+(\uj\dj\ui-\ob{\uj\dj\ui})&=&-\ob{\rho}^{-1}\di\pp+g\widehat\alpha\Tp\kz\,,\\
\dt\Tp+(\uj\dj\Tp-\ob{\uj\dj\Tp})&=&\beta w-(\ob{\rho}\,\ob{c_{\rm p}})^{-1}\di F_i^\prime\,,
\vspace{-1pt}
\label{eq:bappox}
\end{eqnarray}
supplemented by the continuity equation for an incompressible gas, $\dj\uj=0$, 
where $\bu=(u,v,w)$ is the turbulent velocity field, $\rho$ is density, $p$ 
is gas pressure, $g$ is the acceleration due to gravity, $T$ is temperature, 
$c_{\rm p}$ is the specific heat at constant pressure, 
$\widehat\alpha=-\ob{(\partial\ln\rho/\partial\ln T)_{p}}\,/\,\ob T$, 
$F_i$ is the radiative heat flux, $\beta$ is the superadiabatic temperature 
gradient and $\delta_{ij}$ is the Kronecker delta. Primes~($^\prime$) indicate 
Eulerian fluctuations and overbars horizontal averages. These are the starting
equations for the two physical pictures describing the motion of an overturning 
convective eddy, illustrated in Fig.~1.

\noindent
In the first physical picture, adopted by Unno (1967), the turbulent element,
with a characteristic vertical length $\ell$, evolves out of some chaotic 
state and achieves steady motion very quickly. The fluid element maintains 
exact balance between buoyancy force and turbulent drag by continuous exchange 
of momentum with other elements and its surroundings. Thus the acceleration 
terms $\dt\ui$ and $\dt\Tp$ are neglected and the nonlinear advection terms 
provide dissipation (of kinetic energy) that balances the driving terms.
The nonlinear advection terms are approximated by 
$\uj\dj\ui-\ob{\uj\dj\ui}\simeq2w^2/\ell$ and
$\uj\dj\Tp-\ob{\uj\dj\Tp}\simeq2wT^\prime/\ell$. This leads to two 
nonlinear equations which need to be solved numerically together with the 
mean equations of the stellar structure.
%---------------------------------------------------
\figureCoAst{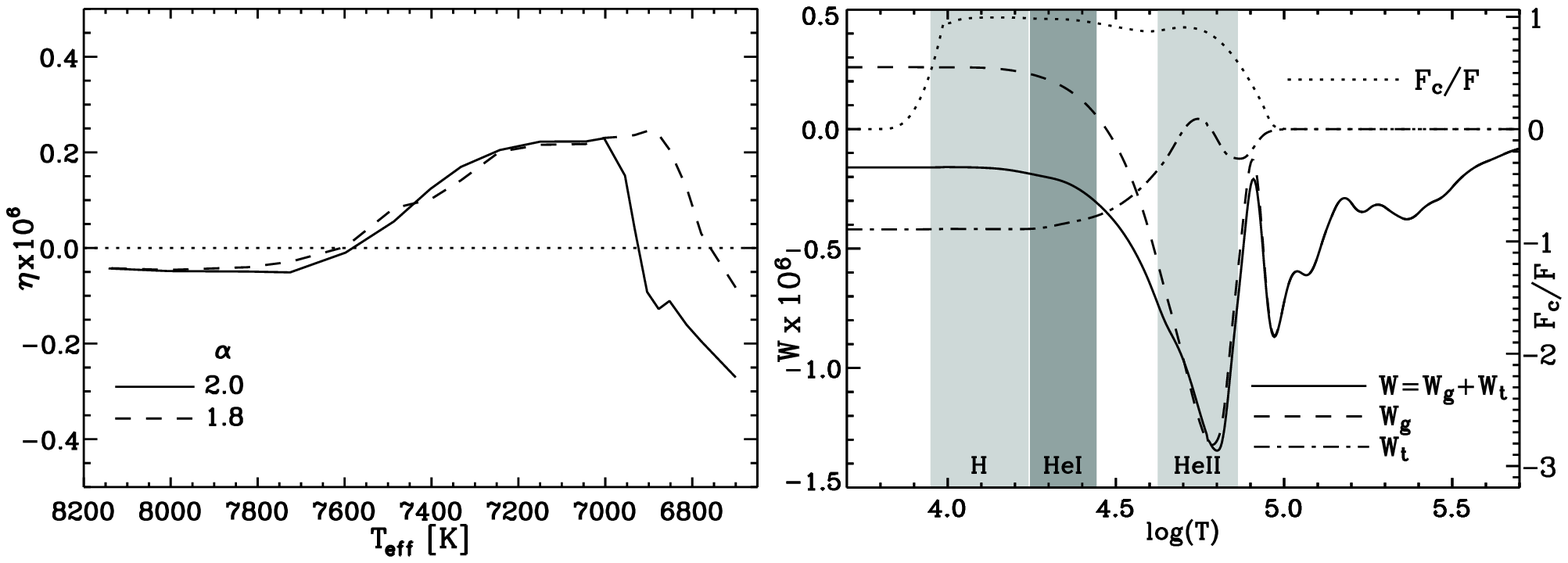}
{Mode stability of an $1.7\,$M$_\odot$ Delta Scuti star computed with
Gough's (1977a,b) convection model. Left: Stability coefficient 
$\eta=\omega_{\rm i}/\omega_{\rm r}$ as a function of surface temperature 
$T_{\rm eff}$ across the IS. Results are shown for the fundamental radial 
mode ($n=1$) and for two values of the mixing-length 
parameter $\alpha$. Positive $\eta$ values indicate mode instability. 
Right: Integrated work integral $W$ as a function of the depth co-ordinate 
$\log(T)$ for a model lying just outside the cool edge 
of the IS ($T_{\rm eff}=6813\,$K). Results are plotted in 
units of $\eta$ and for $\alpha=2.0$. Contributions to $W$ 
(solid curve) arising from the gas pressure perturbation, $W_{\rm g}$ 
(dashed curve), and the turbulent pressure fluctuations, 
$W_{\rm t}$ (dot-dashed curve), are indicated ($W=W_{\rm g}+W_{\rm t}$). 
The dotted curve is the ratio of the convective to the total 
heat flux ${F_{\rm c}/F}$. Ionization zones of H and He 
($5\%$ to $95\%$ ionization) 
are indicated (from Houdek 2000).
\vspace{-9pt}
}
{fig:2}{ht!}{width=\textwidth}

\noindent
The second physical picture, which was generalized by Gough (1965, 1977a,b) to the 
time-dependent case, interprets the turbulent flow by indirect analogy 
with kinetic gas theory. The motion is not steady and one imagines the 
convective element to accelerate from rest followed by an instantaneous 
breakup after the element's lifetime. Thus the nonlinear advection terms are 
neglected in the convective fluctuation equations (1)-(2) but are taken to be
responsible for the creation and destruction of the convective eddies
(Gough 1977a,b). By retaining only the acceleration terms 
the equations become linear with analytical solutions $w\propto\exp(\sigma t)$
and $T^\prime\propto\exp(\sigma t)$ subject to proper periodic spatial boundary 
conditions, where $t$ is time and $\Re(\sigma)$ is the linear convective 
growth rate. The mixing length $\ell$ enters in the calculation of the 
eddy's survival probability, which is proportional to the eddy's internal
shear (rms vorticity), for determining the convective heat and momentum fluxes.
Although the two physical pictures give the same result in a static 
envelope, the results for the fluctuating turbulent 
fluxes in a pulsating star are very different (Gough 1977a). The main 
differences between Unno's and Gough's convection model 
are summarized in Table$\,$1.

\vspace{-10pt}
\section{Application on mode stability in $\delta$ Scuti stars}
\vspace{-3pt}
Fig.~\ref{fig:2} displays the mode stability of an evolving 1.7$\,$M$_\odot$ 
Delta Scuti star crossing the IS. The results were computed with the 
time-dependent, nonlocal convection model by Gough (1977a,b). As demonstrated 
in the right panel of Fig.~\ref{fig:2}, the dominating damping term to the
work integral $W$ for a star located near the red edge is the contribution 
from the turbulent pressure fluctuations $W_{\rm t}$.
%-------------------------------------------------------------------
\figureCoAst{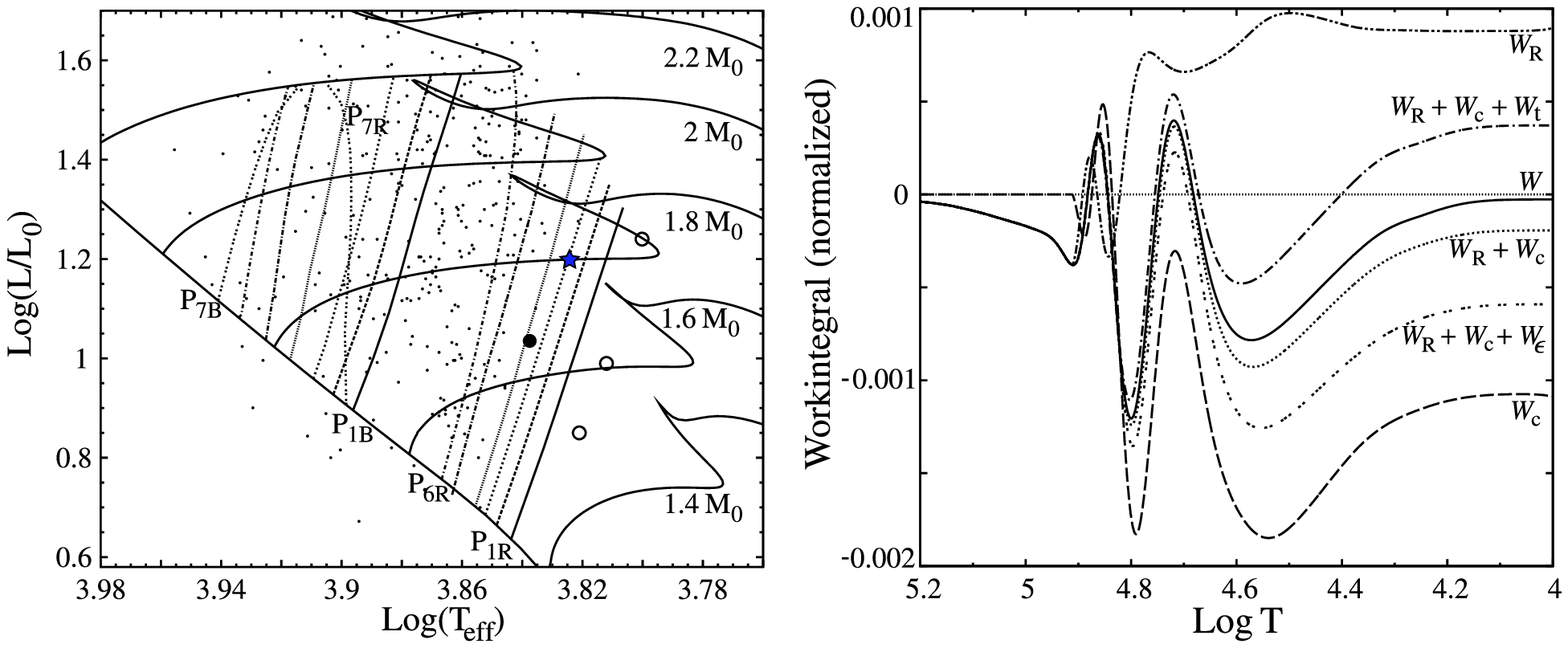}
{Stability computations of Delta Scuti stars which include the viscous
dissipation rate $\epsilon$ of turbulent kinetic energy according 
to Grigahc\`ene et al. (2005).
Left: Blue and red edges of the IS superposed on evolutionary tracks 
on the theorists H-R diagram. The locations of the edges, labelled 
${\rm p}_{n{\rm B}}$ and ${\rm p}_{n{\rm R}}$, are indicated 
for radial modes with orders $1\le n\le7$. Results by Houdek (2000, 
$\alpha=2.0$, see Fig.~\ref{fig:2}) and Xiong \& Deng (2001) for the
gravest p modes are plotted as the filled and open circles respectively. 
Right: Integrated work integral 
$W$ as a function of the depth co-ordinate $\log(T)$ for a stable $n=3$ 
radial mode of a $1.8\,{\rm M}_\odot$ star (see `star' symbol in the 
left panel).  Contributions to $W$ arising from the radiative 
flux, $W_{\rm R}$, the convective flux, $W_{\rm c}$, the turbulent pressure 
fluctuations, $W_{\rm t}$ ($W_{\rm R}+W_{\rm c}+W_{\rm t}$), and from the 
perturbation of the turbulent kinetic energy dissipation, $W_\epsilon$ 
($W_{\rm R}+W_{\rm c}+W_\epsilon$), are indicated 
(adapted from Dupret et al. 2005).
\vspace{-7pt}
}
{fig:3}{ht}{clip,width=\textwidth}

Gabriel (1996) and more recently Grigahc\`ene et al. (2005) generalized
Unno's time-dependent convection model for stability computations of
nonradial oscillation modes. %Furthermore, following Ledoux \& Walraven (1958), 
They included in their mean thermal energy equation 
the viscous dissipation of turbulent kinetic energy, $\epsilon$, 
as an additional heat source.  The dissipation of turbulent kinetic energy
is introduced in the conservation equation for the turbulent kinetic energy
$K:=\ob{\ui\ui}/2$ (e.g. Tennekes \& Lumley 1972,$\,\mathsection$3.4; 
Canuto 1992; \hbox{Houdek \& Gough 1999)}:
\begin{equation}
\vspace{-3pt}
{\rm D}_tK+\dj(\ob{K\uj}+\ob{\rho}^{-1}\ob{p^\prime\uj})-\nu\partial^2_iK=
-\ob{\ui\uj}\dj U_i+g\widehat\alpha\ob{\uj\Tp}-\epsilon\,,
%\vspace{-3pt}
\label{eq:tke}
\end{equation}
where ${\rm D}_t$ is the material derivative, $U_i$ is the average (oscillation)
velocity, i.e. the total velocity $\tilde\ui=U_i+\ui$, and $\nu$ is the constant
kinematic viscosity (in the limit of high Reynolds numbers the molecular 
transport term can be neglected).
The first and second term on the right of Eq.~(\ref{eq:tke}) are the 
shear and buoyant productions of turbulent kinetic energy, whereas the last term
$\epsilon=\nu\ob{(\dj\ui+\di\uj)^2}/2$ is the viscous dissipation of turbulent 
kinetic energy into heat. This term is also present in the mean thermal
energy equation, but with opposite sign. The linearized perturbed mean thermal
energy equation for a star pulsating radially with complex angular frequency
$\omega=\omega_{\rm r}+{\rm i}\omega_{\rm i}$ can then be written, in the 
absence of nuclear reactions, as
(`$\delta$' denotes a Lagrangian fluctuation and I omit overbars in 
the mean quantities):
\begin{equation}
{\rm d}\delta L/{\rm d}m=-{\rm i}\omega c_{\rm p}T({\delta T}/{T}- 
\nabla_{\rm ad}{\delta p}/{p})+\delta\epsilon\,,
\label{eq:peq}
\end{equation}
where $m$ is the radial mass co-ordinate, 
$\nabla_{\rm ad}=(\partial\ln T/\partial\ln p)_s$ and $L$ is the total 
(radiative and convective) luminosity. 
Grigahc\`ene et al. (2005) evaluated $\epsilon$ from a turbulent kinetic 
energy equation which was derived without the assumption of the Boussinesq 
approximation. Furthermore it is not obvious whether the dominant buoyancy 
production term, $g\widehat\alpha\ob{\uj\Tp}$ (see Eq.~\ref{eq:tke}), was 
included in their turbulent kinetic energy equation and so
\hbox{in their expression for $\epsilon$.}

\noindent
Dupret et al. (2005) applied the convection model of Grigahc\`ene et al. (2005)
to Delta Scuti and $\gamma$ Doradus stars and reported well defined red edges. 
The results of their stability analysis for Delta Scuti stars are 
depicted in Fig.~\ref{fig:3}. The left panel compares the location of the
red edge with results reported by Houdek (2000, see also Fig.~\ref{fig:2}) and 
Xiong \& Deng (2001). The right panel of Fig.~\ref{fig:3} displays the 
individual contributions to the accumulated work integral $W$ for a star 
located near the red edge of the $n=3$ mode (indicated by the `star' symbol 
in the left panel). It demonstrates the near cancellation effect between 
the contributions of the turbulent kinetic energy dissipation , $W_\epsilon$, 
and turbulent pressure, $W_{\rm t}$, making the contribution from the 
fluctuating convective heat flux, $W_{\rm c}$, the dominating damping term.
The near cancellation effect between $W_\epsilon$ and $W_{\rm t}$ was 
demonstrated first by Ledoux \& Walraven (1958, $\mathsection$65) 
(see also Gabriel 1996) by writing the sum of both work integrals as:
\begin{equation}
\vspace{-2pt}
W_\epsilon+W_{\rm t}=3\pi/2\int_{m_{\rm b}}^M(5/3-\gamma_3)
\Im(\delta p_{\rm t}^*\delta\rho)\rho^{-2}\,{\rm d}m\,,
\vspace{-2pt}
\label{eq:wi}
\end{equation}
where $M$ is the stellar mass, $m_{\rm b}$ is the enclosed mass at the bottom 
of the envelope and $\gamma_3\equiv1\!+\!(\partial\ln T/\partial\ln\rho)_s$ 
($s$ is specific entropy) is the third adiabatic exponent. Except in ionization zones 
$\gamma_3\simeq 5/3$ and consequently $W_\epsilon+W_{\rm t}\simeq0$.
%------------------------------------------
\figureCoAst{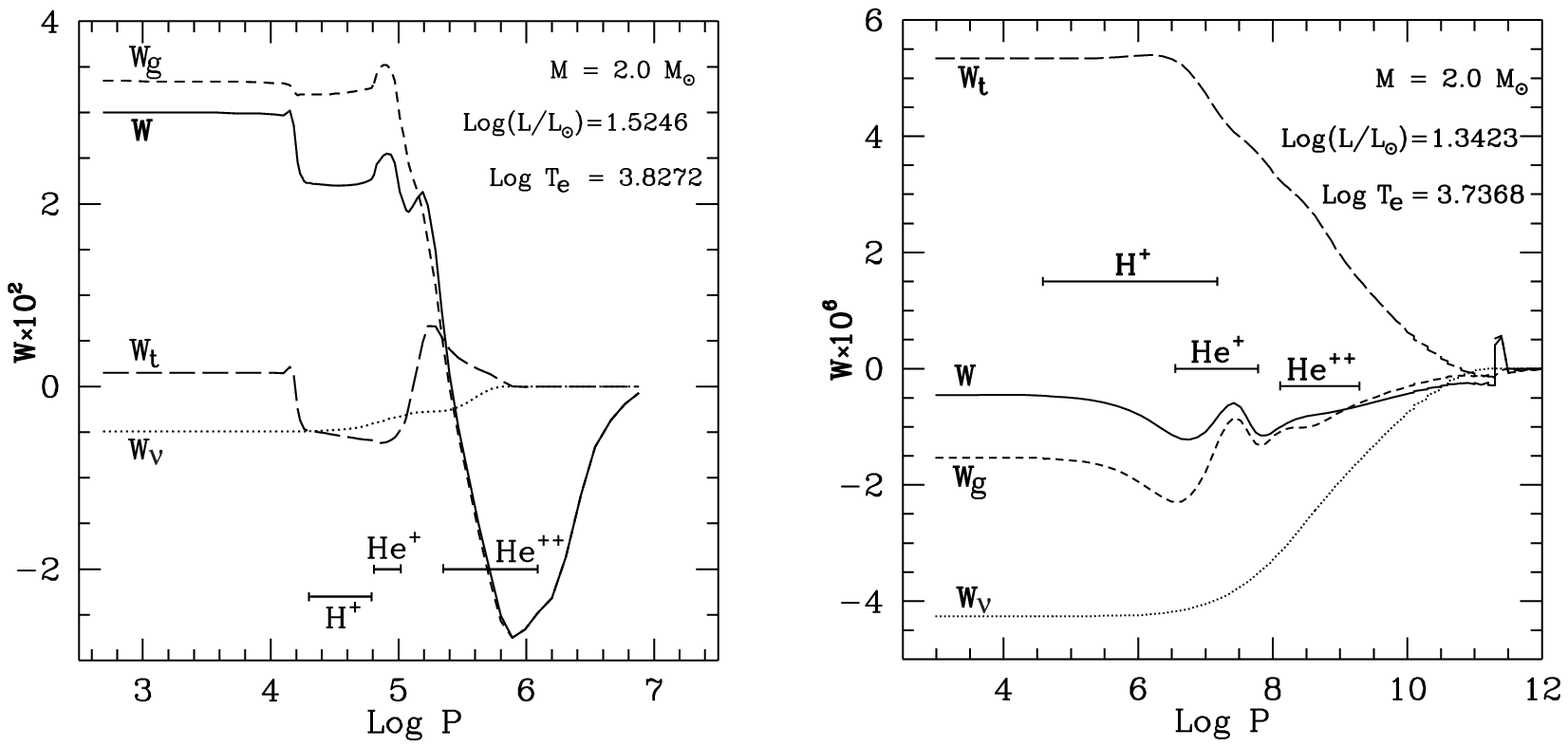}
{Accumulated work integral $W$ as a function of the depth co-ordinate $\log(p)$.
Results are shown for the $n=1$ radial mode of a Delta Scuti star located 
inside the IS (left panel) and outside the red edge of the IS (right panel). 
The stability calculations include viscous dissipation by the 
small-scale turbulence (Xiong 1989; see Eq.~\ref{eq:Wnu}). Contributions
to $W$ (solid curve) arising from the fluctuating gas pressure, $W_{\rm g}$ 
(dashed curve), the turbulent pressure perturbations, $W_{\rm t}$ 
(long-dashed curve), and from the turbulent viscosity, $W_\nu$ (dotted curve),
are indicated ($W=W_{\rm g}+W_{\rm t}+W_\nu$). The ionization zones of H and 
He are marked (adapted from Xiong \& Deng 2007).
\vspace{-8pt}
}
{fig:4}{ht}{clip,width=0.95\textwidth}

The convection model by Xiong (1977, 1989) uses transport equations for the 
second-order moments of the convective fluctuations. 
In the transport equation for the turbulent kinetic energy Xiong adopts the
approximation by Hinze (1975) for the turbulent dissipation rate, 
i.e. $\epsilon=2\chi k(\ob{\ui\ui}\rho^2/3\ob{\rho^2})^{3/2}$,
where $\chi=0.45$ is the Heisenberg eddy coupling coefficient and 
$k\propto\ell^{-1}$ is the wavenumber of the energy-containing eddies.
However, Xiong does not provide a work integral for $\epsilon$ 
(neither does Unno et al. 1989, $\mathsection$26,30) but includes the viscous
damping effect of the small-scale turbulence in his model.
The convection models considered here describe only the largest, most 
energy-containing eddies and ignore the dynamics of the small-scale eddies 
lying further down the turbulent cascade. Small-scale turbulence does, however, 
contribute directly to the turbulent fluxes and, under the assumption 
that they evolve isotropically, they generate an effective viscosity 
$\nu_{\rm t}$ which is felt by a particular pulsation mode as an 
additional damping effect. The turbulent viscosity can be estimated as
(e.g. Gough 1977b; Unno$\,$et$\,$al.$\,$1989, $\mathsection$20) 
$\nu_{\rm t}\simeq\lambda(\ob{ww})^{1/2}\ell$, where $\lambda$ is a parameter
of order unity. The associated work integral $W_\nu$ can be written in Cartesian
co-ordinates as (Ledoux \& Walraven 1958, $\mathsection$63) 
%;Unno$\,$et$\,$al.$\,$1989, $\mathsection$30)
\begin{equation}
\vspace{-2pt}
W_\nu=-2\pi\;\omega_{\rm r}\int_{m_{\rm b}}^M\nu_{\rm t}
\left[e_{ij}e_{ij}-\frac{1}{3}\left(\nabla\cdot\bx\right)^2\right]\,{\rm d}m\,,
%\vspace{-2pt}
\label{eq:Wnu}
\end{equation}
where $e_{ij}=(\dj\xi_i+\di\xi_j)/2$ and $\bx$ is the displacement eigenfunction.
Xiong \& Deng (2001, 2007) modelled successfully the IS of 
Delta Scuti and red giant stars and found the dominating damping effect 
to be the turbulent viscosity (Eq. \ref{eq:Wnu}). This is illustrated in
Fig.~\ref{fig:4} for two Delta Scuti stars: one is located inside 
the IS (left panel), the other outside the cool edge of the IS (right panel).
The contribution from the small-scale turbulence was also the 
dominant damping effect in the stability calculations by 
Xiong et al. (2000) of radial p modes in the Sun, although the
authors still found unstable modes with orders between $11\le n\le23$. 
The importance of the turbulent damping was reported first by
Goldreich \& Keeley (1977) and later by Goldreich \& Kumar (1991), who found all
solar modes to be stable only if turbulent damping was included in their
stability computations.
In contrast, Balmforth (1992), who adopted the convection model of 
Gough (1977a,b), found all solar p modes to be stable due mainly to the 
damping of the turbulent pressure perturbations, $W_{\rm t}$, and 
reported that viscous damping, $W_\nu$, is about one order of magnitude 
smaller than the contribution of $W_{\rm t}$. Turbulent viscosity 
(Eq.~\ref{eq:Wnu}) leads always to mode damping, where as the perturbation 
of the turbulent kinetic energy dissipation, $\delta\epsilon$ 
(see Eq. \ref{eq:peq}), can contribute to both damping and driving of the 
pulsations (Gabriel 1996). The driving effect of 
$\delta\epsilon$ was shown by Dupret et al. (2005) 
for a $\gamma$ Doradus star.
\vspace{-5pt}
\section{Summary}
\vspace{-3pt}
We discussed three different mode stability calculations of Delta Scuti stars 
which successfully reproduced the red edge of the IS. Each of these computations
adopted a different time-dependent convection description.
The results were discussed by comparing work integrals. All convection 
descriptions include, although in different ways, the perturbations of 
the turbulent fluxes. Gough (1977a), Xiong (1977, 1989), and 
Unno$\,$et$\,$al.$\,$(1989) did not include the contribution $W_\epsilon$ 
to the work integral because in the Boussinesq approximation 
(Spiegel \& Veronis 1960) the viscous dissipation is neglected in the 
thermal energy equation. In practise, however, this term may be important.
Grigahc\`ene et al. (2005) included $W_\epsilon$ but ignored the damping 
contribution of the small-scale turbulence $W_\nu$, which 
was found by Xiong \& Deng (2001, 2007) to be the dominating damping term. 
The small-scale damping effect was also ignored in the calculations by 
Houdek (2000). A more detailed comparison of the convection descriptions has 
not yet been made but Houdek \& Dupret have begun to address this problem.

\vspace{-5pt}
\acknowledgments{
I am grateful to Douglas Gough for helpful discussions
and to the HelAs organizing committee for travel support.
Support by the UK Science and Technology Facilities Council is
acknowledged.}

\vspace{-5pt}
\References{
\vspace{-2pt}
\rfr Baker N., Gough D.O. 1979, ApJ, 234, 232
\rfr Balmforth N.J. 1992, MNRAS, 255, 603
%\rfr Balmforth M.J., Gough D.O. 1988, in Seismology of the Sun and Sun-Like Stars,
%     Domingo V., Rolfe E.J., eds, ESA SP-286, Noordwijk, p.$\,$47
\rfr Bono G., Caputo F., Castellani V., et al. 1995, ApJ, 442, 159
\rfr Bono G., Marconi M., Stellingwerf R. 1999, ApJSS, 122, 167
\rfr Canuto V. 1992, ApJ, 392, 218
%\rfr Cheng Q.L., Xiong D.R. 1997, A\&A, 319, 981
%\rfr Dupree R.G. 1977a, ApJ, 211, 509
%\rfr Dupree R.G. 1977b, ApJ, 214, 502
\rfr Dupree R.G. 1977, ApJ, 215, 232
%\rfr Dupret M.-A., Grigahc\`ene A., Garrido R., Gabriel M., Scuflaire R. 2005,
\rfr Dupret M.-A., Grigahc\`ene A., Garrido R., et al. 2005, A\&A, 435, 927
\rfr Gabriel M. 1996, Bull. Astron. Soc. India, 24, 233
\rfr Goldreich P., Keeley D.A. 1977, ApJ, 211, 934
\rfr Goldreich P., Kumar P. 1991, ApJ, 374, 366
\rfr Gonzi G. 1982, A\&A, 110, 1
\rfr Gough D.O. 1965, in Geophys. Fluid Dynamics, Woods Hole Oceanographic
     Institutions, Vol.$\,$2, Woods Hole, Mass., p.$\,$49
\rfr Gough D.O. 1969, J. Atmos. Sci., 26, 448
\rfr Gough D.O. 1977a, ApJ, 214, 196
\rfr Gough D.O. 1977b, in Problems of Stellar Convection,
     Spiegel E.A., Zahn J.-P., eds, Springer-Verlag, Berlin, p.$\,$15
%\rfr Gough D.O. 1978, in Proc. Workshop on Solar Rotation, Belvedere G.,
%     Patern\`o L., eds, Catania University Press, Catania, p.$\,$337
%\rfr Grigahc\`ene A., Dupret M.-A., Gabriel M., Garrido R., Scuflaire R. 2005,
\rfr Grigahc\`ene A., Dupret M.-A., Gabriel M., et al. 2005, A\&A, 434, 1055
%\rfr Grossman S. 1996, MNRAS, 279, 305
%\rfr Grossman S., Narayan R., Arnett D. 1993, ApJ, 407, 284 
\rfr Hinze J.O. 1975, Turbulence, McGraw-Hill, New York
\rfr Houdek G. 1997, in Proc. IAU Symp. 181: Sounding Solar and Stellar
     Interiors, Schmider F.-X., Provost J., eds, Nice Observatory, p.$\,$227
\rfr Houdek G. 2000, in Delta Scuti and Related Stars, Breger M., 
     Montgomery M.H., eds, ASP Conf. Ser., Vol. 210, Astron. Soc. Pac., 
     San Francisco, p.$\,$454 
\rfr Houdek G., Gough D.O. 1999, in Theory \& Tests of Convection in Stellar
     Structure, Guinan E.F., Montesinos B., eds, ASP Conf. Ser. 173,
     San Francisco, p.$\,$237
%\rfr Houdek G., Balmforth N.J., Christensen-Dalsgaard J., Gough D.O. 1999
%     A\&A, 351, 582
\rfr Kuhfu\ss\ R. 1986, A\&A, 160, 116
%\rfr Kundu P.K. 1990, Fluid Mechanics, Academic Press, San Diego
\rfr Ledoux P., Walraven T. 1958, in Handbuch der Physik, Vol. LI, 
     Fl\"ugge S., ed., Springer-Verlag, Berlin, p.$\,$353
\rfr Schatzman E. 1956, Annales d'Astrophysique, 19, 51
\rfr Spiegel E.A., Veronis G. 1960, ApJ, 131, 442 (correction: ApJ, 135, 665)
\rfr Stellingwerf R.F. 1982, ApJ, 262, 330
%\rfr Stellingwerf R.F. 1984a, ApJ 277, 322
\rfr Stellingwerf R.F. 1984, ApJ, 284, 712
\rfr Tennekes H., Lumley J.L. 1972, A First Course in Turbulence, The MIT Press
\rfr Townsend A.A. 1976, The Structure of Turbulent Shear Flow, CUP  
\rfr Unno W. 1967, PASJ, 19, 40
\rfr Unno W. 1977, in Problems of Stellar Convection,
     Spiegel E.A., Zahn J.-P., eds, Springer-Verlag, Berlin, p.$\,$315
\rfr Unno W., Osaki Y., Ando H., Saio H, Shibahashi H. 1989, Nonradial 
     Oscillations of Stars, Second Edition, University of Tokyo Press
\rfr Xiong D.R. 1977, Acta Astron. Sinica, 18, 86
\rfr Xiong D.R. 1989, A\&A, 209, 126
%\rfr Xiong D.R., Deng L., Cheng Q.L. 1998, ApJ, 500, 449
\rfr Xiong D.R., Cheng Q.L., Deng L. 2000, MNRAS, 319, 1079
\rfr Xiong D.R., Deng L. 2001, MNRAS, 324, 243
\rfr Xiong D.R., Deng L. 2007, MNRAS, 378, 1270
}%\References

\vspace{-8pt}
\begin{center}
\section{Discussion}
\end{center}
\vspace{-2pt}
\discuss{Christensen-Dalsgaard}{How does the mixing length affect the 
red edge of the $\gamma\;$Dor instability strip?}
\discuss{Houdek}{The location of the red edge is predominantly determined 
by radiative damping which gradually dominates over the driving effect of the
so-called convective flux blocking mechanism (Dupret et al. 2005). A change 
in the mixing length will not only affect the depth of the envelope 
convection zone but also the characteristic time scale of the convection 
and consequently the stability of g modes with different pulsation periods.
A calibration of the mixing length to match the observed location of the
$\gamma\;$Dor instability strip will also calibrate the depth of the
convection zone at a given surface temperature.}

\end{document}